\documentclass[slac_one]{revtex4}
\usepackage{graphicx}
\usepackage{subfigure}
\usepackage{fancyhdr}
\begin{document}

\title{Tau Reconstruction and Identification Performance at ATLAS} 

%

\author{Bj\"orn Gosdzik\\on behalf of the ATLAS Collaboration}
\affiliation{Deutsches Elektronen-Synchrotron, Notkestrasse 85, D-22607 Hamburg, GERMANY}
%

\begin{abstract}
For many signals in the Standard Model including the Higgs boson, and for new physics like Supersymmetry, $\tau$ leptons represent an important signature. This work shows the performance of the ATLAS $\tau$ reconstruction and identification algorithms. It will present a set of studies based on data taken in 2010 at a center-of-mass energy of $\sqrt{s}$ = 7 TeV. We measured some of the basic input quantities used for these identification methods from selected reconstructed $\tau$ candidates and compared the results to the prediction of different Monte Carlo simulation models. For early data taking a cut-based identification method will be used. We also measured the background efficiency for the cut-based $\tau$ identification.
\end{abstract}

\maketitle

\thispagestyle{fancy}


\section{Introduction}\label{sec:intro}

In the Standard Model, a large number of $\tau$ leptons is expected from the decay of $Z$ and $W$ bosons with 100 $\rm pb^{-1}$ of data. $\tau$ leptons also play an important role in searches for new phenoma like the Standard Model Higgs boson, MSSM Higgs bosons and SUSY with large tan $\beta$ since $\tau$ leptons can differentiate between SUSY models based on polarization information. They have a mass of $\rm m_{\tau}$ = 1.78 GeV and decay $\approx$ 65\% of the time hadronically.

$\tau$ leptons in ATLAS typically have a collimated calorimetric cluster, 1 or 3 charged decay products and a displaced secondary vertex in the case of 3-prong decays.

\section{Reconstruction and Identification}
\begin{figure}[htb]
\centering
\subfigure{
  \includegraphics[width=0.40\textwidth]{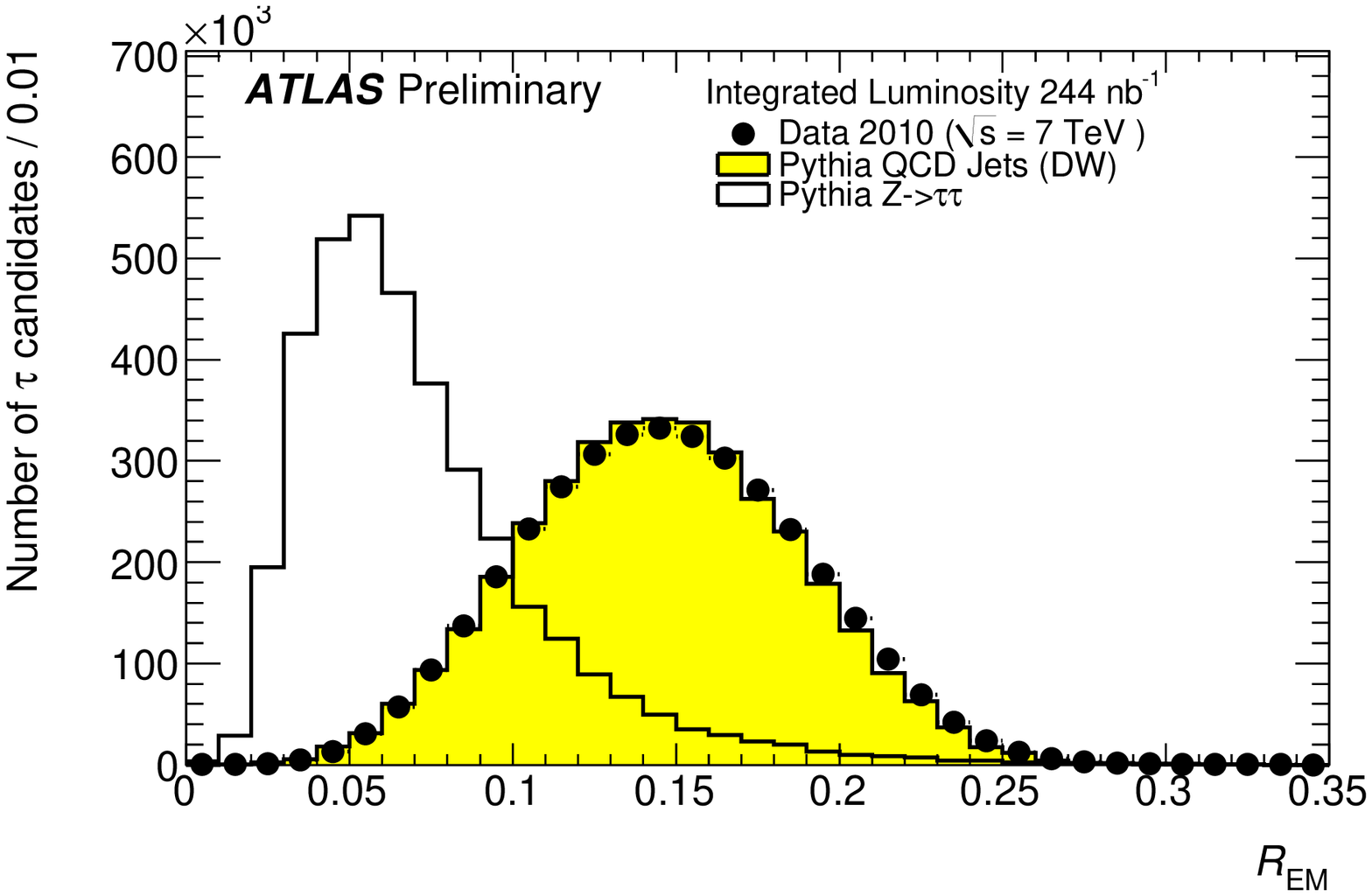}
  \label{emradius}
}
\subfigure{
  \includegraphics[width=0.40\textwidth]{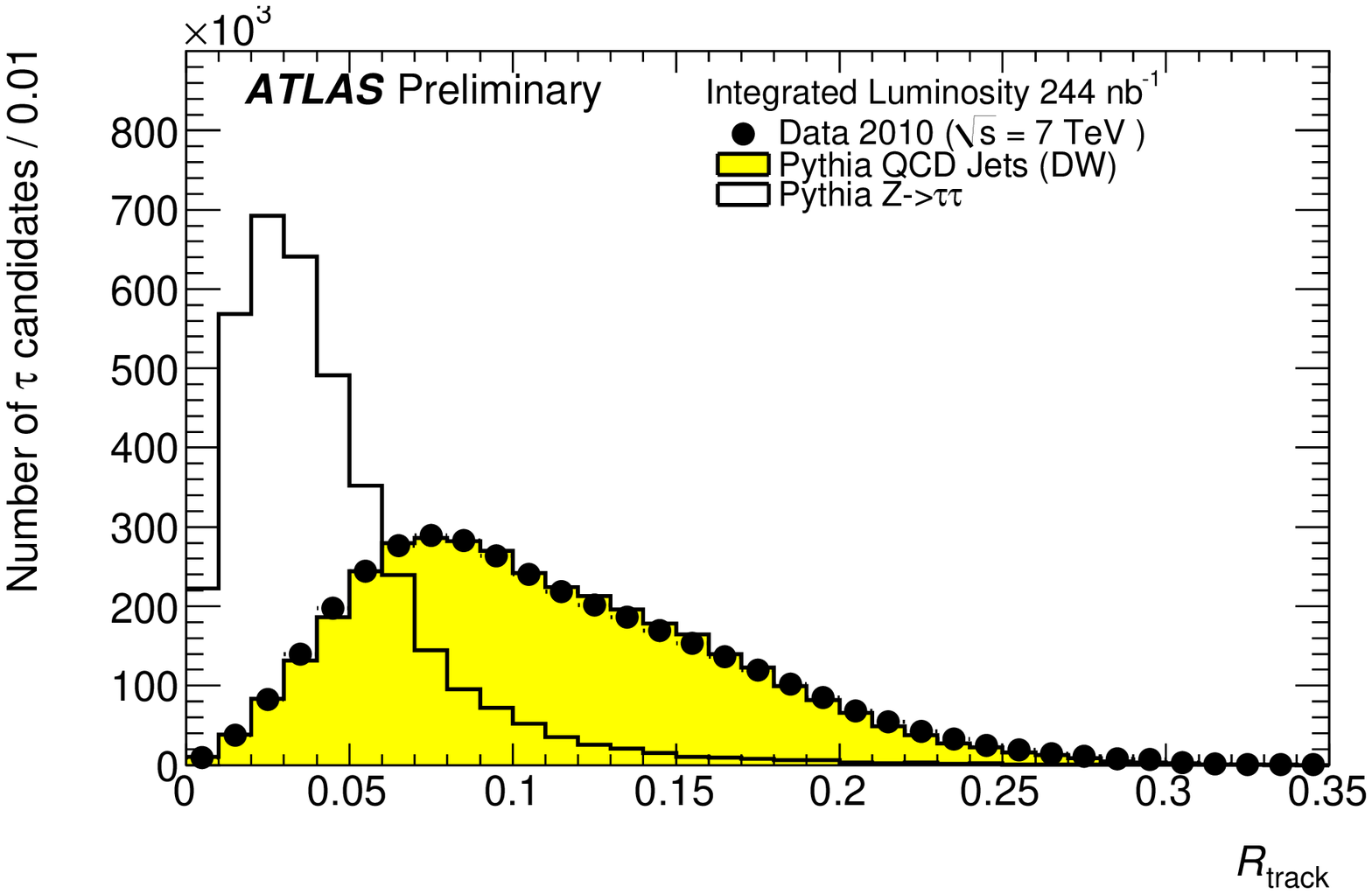}
  \label{trackradius}
}
\caption{EM radius (left) and track radius (right), see section~\ref{sec:bkg} for definitions \protect\cite{conf086}.}
\label{fig:var}
\end{figure}
The studies are based on data collected with the ATLAS detector~\cite{ATLAS} at a center-of-mass energy of $\rm \sqrt{s}$ = 7 TeV and correspond to an integrated luminosity of approximately $\mathcal L$ = 244 $\rm nb^{-1}$. Dedicated cuts on the data, haven been applied to select events with back-to-back jets and to enrich the sample with fake $\tau$ candidates from QCD processes that form the primary background~\cite{conf086} in searches with $\tau$ lepton final states.

To compare the distribution of the variables used for the $\tau$ reconstruction and identification we used predictions from QCD jets Monte Carlo (MC) samples, generated with the Pythia DW tune~\cite{tev4lhc}. The reconstruction of hadronically decaying $\tau$ leptons starts from either calorimeter or track seeds:
\begin{itemize}
\item Track-seeded candidates start with a seeding track of $p_\mathrm{T} >$ 6 GeV, $|\eta|\ <$ 2.5, and satisfy quality criteria on the impact parameter with respect to the interaction vertex ($|d_{0}|\ <$ 2 mm and $|z_{0}| \times$ sin$\theta\ <$ 10 mm).
\item Calorimeter-seeded candidates consist of calorimeter jets reconstructed with the anti-Kt algorithm (using a distance parameter $D$ = 0.4) starting from topological clusters with a calibrated $E_\mathrm{T} >$ 10 GeV and $|\eta|\ <$ 2.5.
\item Candidates are labeled double-seeded when a seed track and a seed jet are within a distance of \mbox{$\rm \Delta R=\sqrt{(\Delta \eta)^2 + (\Delta\phi)^2} <$ 0.2.}
\end{itemize}
Seven variables are currently used as inputs for the identification algorithms to distinguish $\tau$ leptons from QCD jets. The variables electromagnetic (EM) radius and track radius are shown in Fig.~\ref{fig:var}.


\section{Background Rejection in QCD events}\label{sec:bkg}
\begin{figure}[htb]
\centering
\subfigure{
  \includegraphics[width=0.40\textwidth]{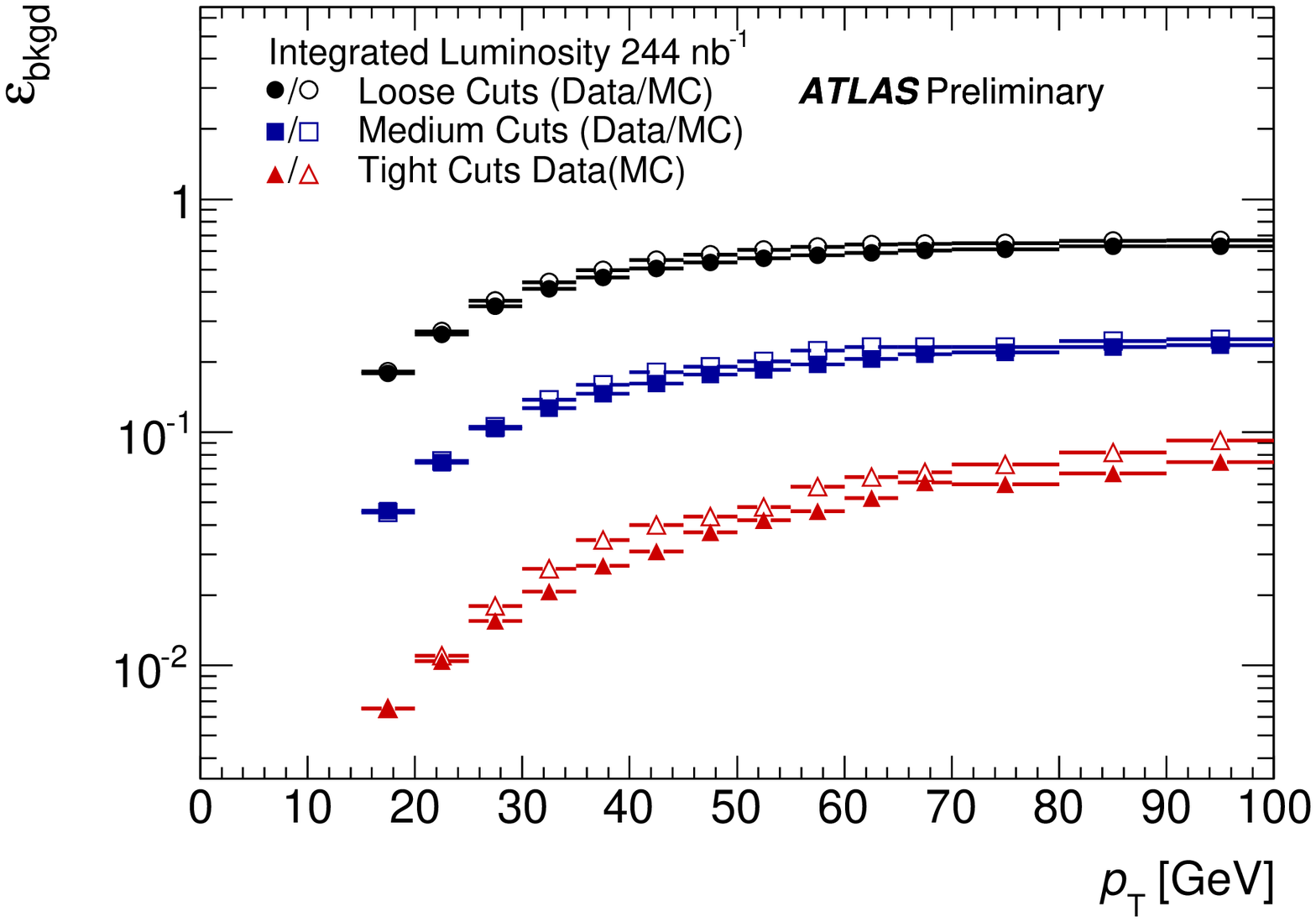}
  \label{subfig:bkg}
}
\subfigure{
  \includegraphics[width=0.40\textwidth]{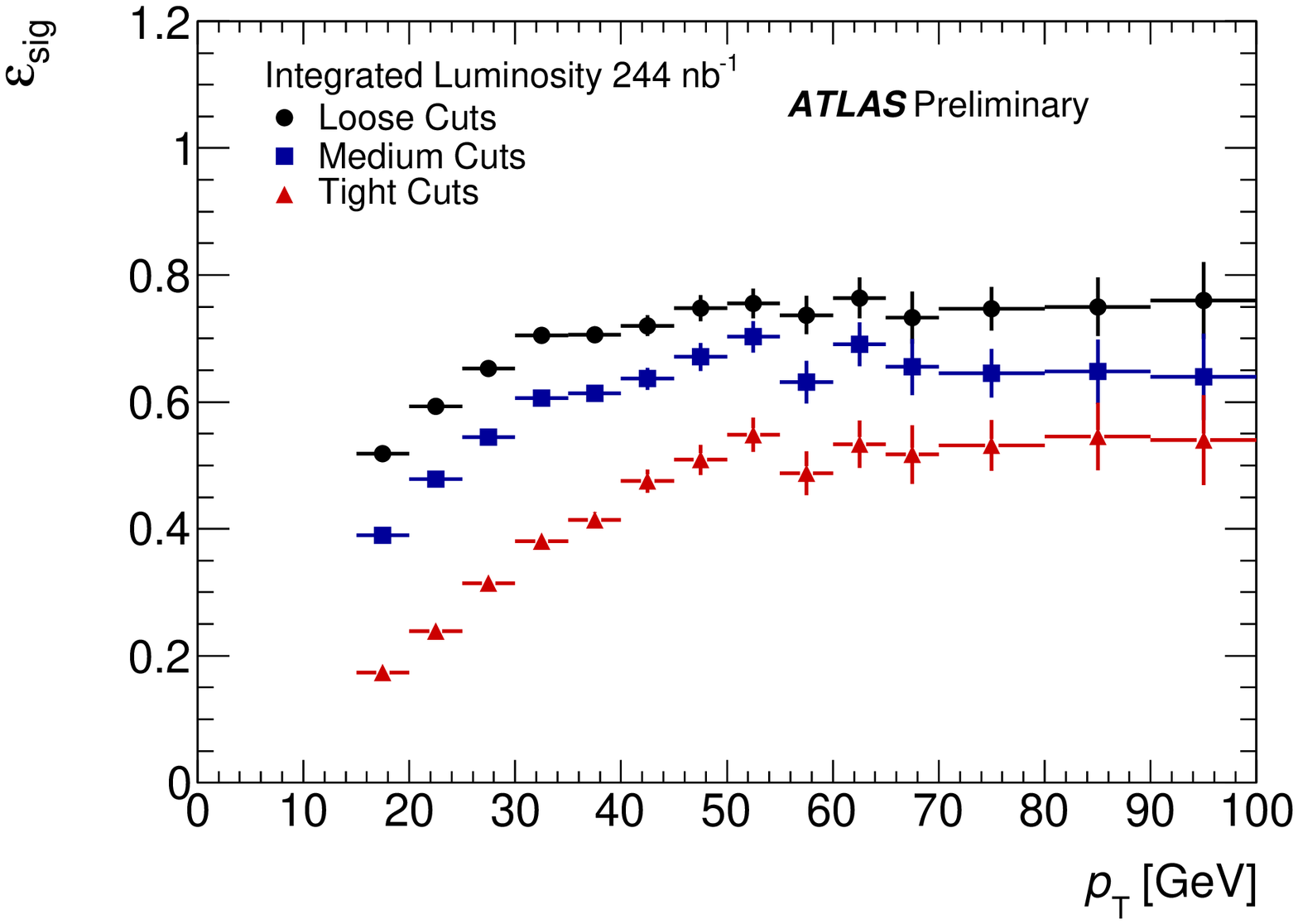}
  \label{subfig:sig}
}
\caption{Background efficiency from Data/MC as a function of the reconstructed $p_\mathrm{T}$ of the $\tau$ candidate (left); signal efficiency from $Z \to \tau\tau$ MC as a function of the reconstructed $p_\mathrm{T}$ (right) \protect\cite{conf086}.}
\label{fig:eff}
\end{figure}
To identify $\tau$ leptons after the reconstruction, three independent identification (ID) algorithms were studied: simple cuts, boosted decision trees (BDT) and a projective likelihood (LL).

\begin{figure}[htb]
\centering
\subfigure{
  \includegraphics[width=0.40\textwidth]{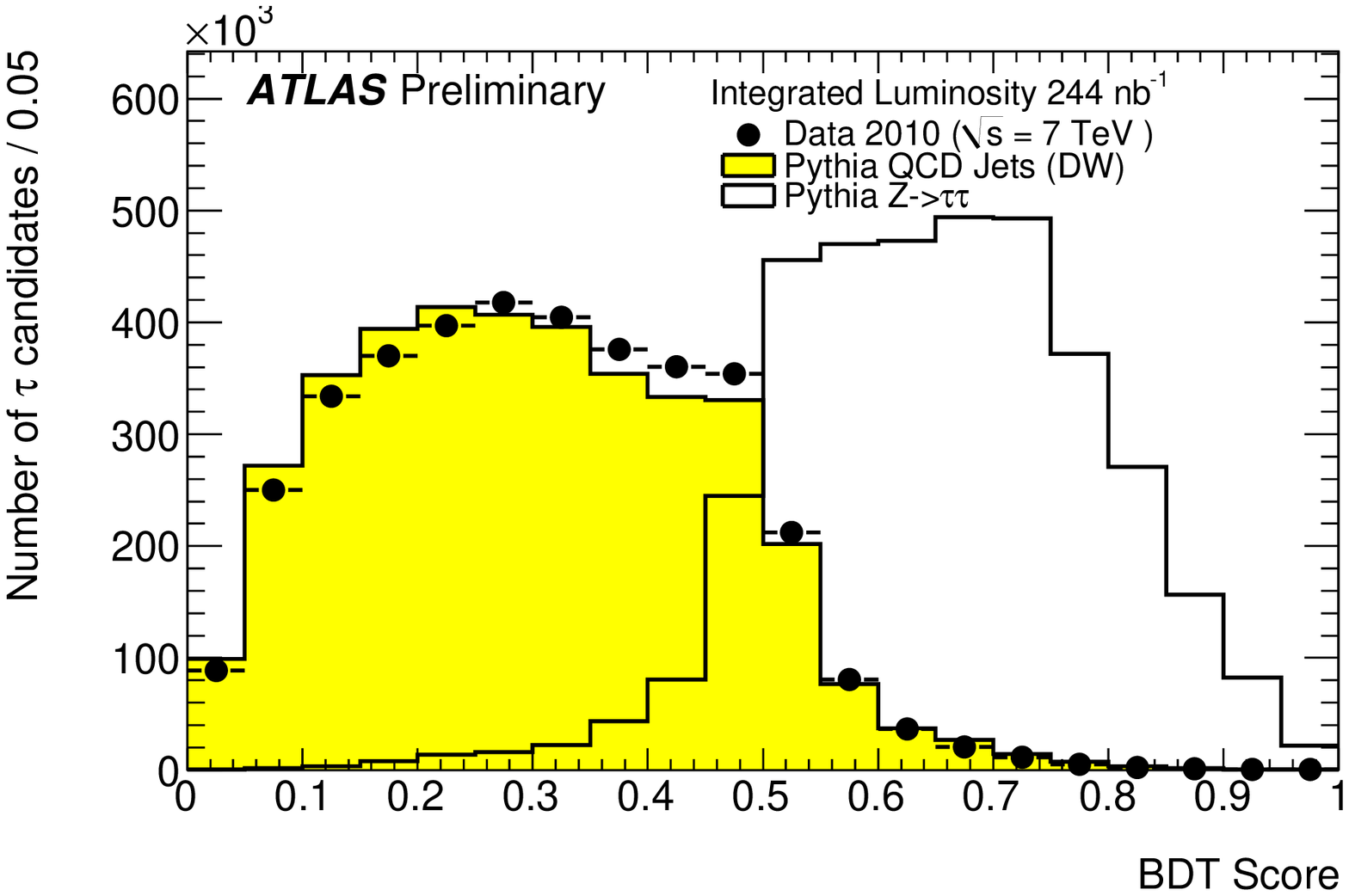}
  \label{subfig:bdt}
}
\subfigure{
  \includegraphics[width=0.40\textwidth]{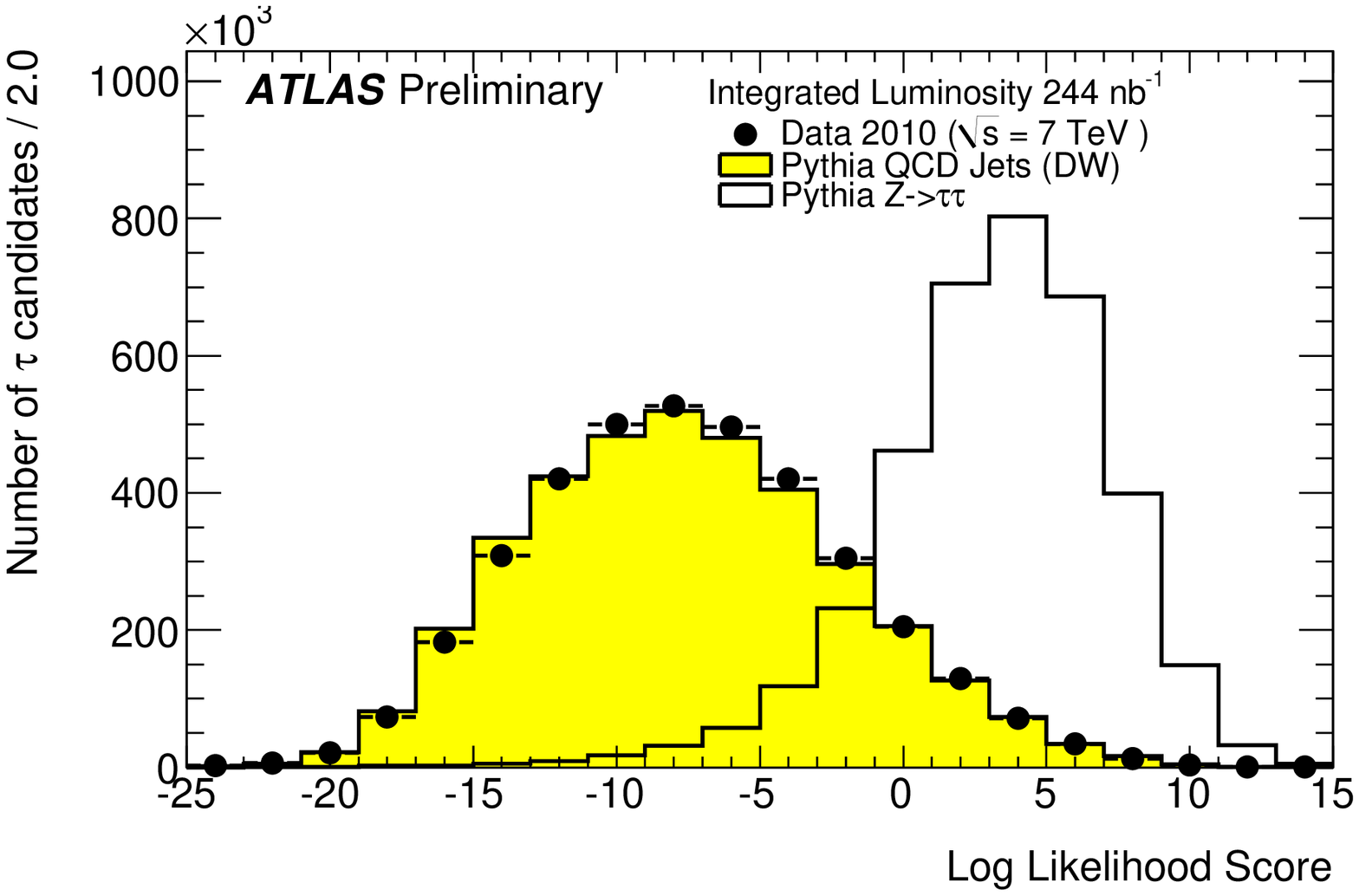}
  \label{subfig:llh}
}
\caption{BDT jet score (left) and LL score (right) for $\tau$ candidates in data and MC samples \protect\cite{conf086}.}
\label{fig:id}
\end{figure}

The cut-based ID uses three variables: $R_\mathrm{EM} =  \frac{\sum_{i}^{\Delta R_i< 0.4}E_{\mathrm{T},i}^{\mathrm{EM}}\Delta R_i}{\sum_{i}^{\Delta R_i< 0.4}E_{\mathrm{T},i}^{\mathrm{EM}}}$, $R_\mathrm{track}  =  \frac{\sum_{i}^{\Delta R_i< 0.2}p_{\mathrm{T},i}\Delta R_i}{\sum_{i}^{\Delta R_i< 0.2}p_{\mathrm{T},i}}$ and $f_\mathrm{trk,1} = \frac{p_{\mathrm{T, 1}}^{\mathrm{track}}}{p_{\mathrm{T}}^{\tau,vis}}$. Three selections corresponding to signal efficiencies of 30\% (tight), 50\% (medium), and 60\% (loose) are optimized to maximize the rejection of QCD jets. The background efficiency for all three selections is shown in Table~\ref{tab:cut_eff_bkgd}. The background efficiency $\varepsilon^{\prime}_\mathrm{bkgd}$ requires $\rm n_{track}$ = 1 or $\rm n_{track}$ = 3. Figure~\ref{fig:eff} shows the background efficiency from Data/MC and signal efficiency from $Z \to \tau\tau$ MC for the cut-based ID. Figure~\ref{fig:id} (left) shows the BDT jet score and Fig.~\ref{fig:id} (right) the likelihood score. The number of $\tau$ candidates in the MC samples is normalized to the number of $\tau$ candidates in the data. Very good agreement between the data and the prediction of QCD MC is observed.
\begin{table}[htbp]
    \centering
    \begin{tabular}{l | c c | c c}
        \hline\hline  
        Selection    &  $\varepsilon_\mathrm{bkgd}$ (data) & $\varepsilon_\mathrm{bkgd}$ (MC) 
                         &  $\varepsilon^{\prime}_\mathrm{bkgd}$ (data) & $\varepsilon^{\prime}_\mathrm{bkgd}$ (MC) \\
        \hline
        loose   &  $(3.2 \pm 0.2) \times 10^{-1}$ & $3.4 \times 10^{-1}$ & $(9.4 \pm 0.6) \times 10^{-2}$ & $10 \times 10^{-2}$ \\
        medium  &  $(9.5 \pm 1.0) \times 10^{-2}$ & $9.9 \times 10^{-2}$ & $(3.1 \pm 0.4) \times 10^{-2}$ & $3.3 \times 10^{-2}$ \\
        tight   &  $(1.6 \pm 0.3) \times 10^{-2}$ & $1.9 \times 10^{-2}$ & $(5.6 \pm 0.9) \times 10^{-3}$ & $6.8 \times 10^{-3}$ \\        
       \hline\hline
    \end{tabular}
    \caption{Background efficiencies for loose, medium, and tight selection cuts.  The measured
             background efficiencies in data are compared to the MC DW tune prediction \protect\cite{conf086}.}
    \label{tab:cut_eff_bkgd}
\end{table}

\begin{figure}[htb]
\centering
\subfigure{
  \includegraphics[width=0.40\textwidth]{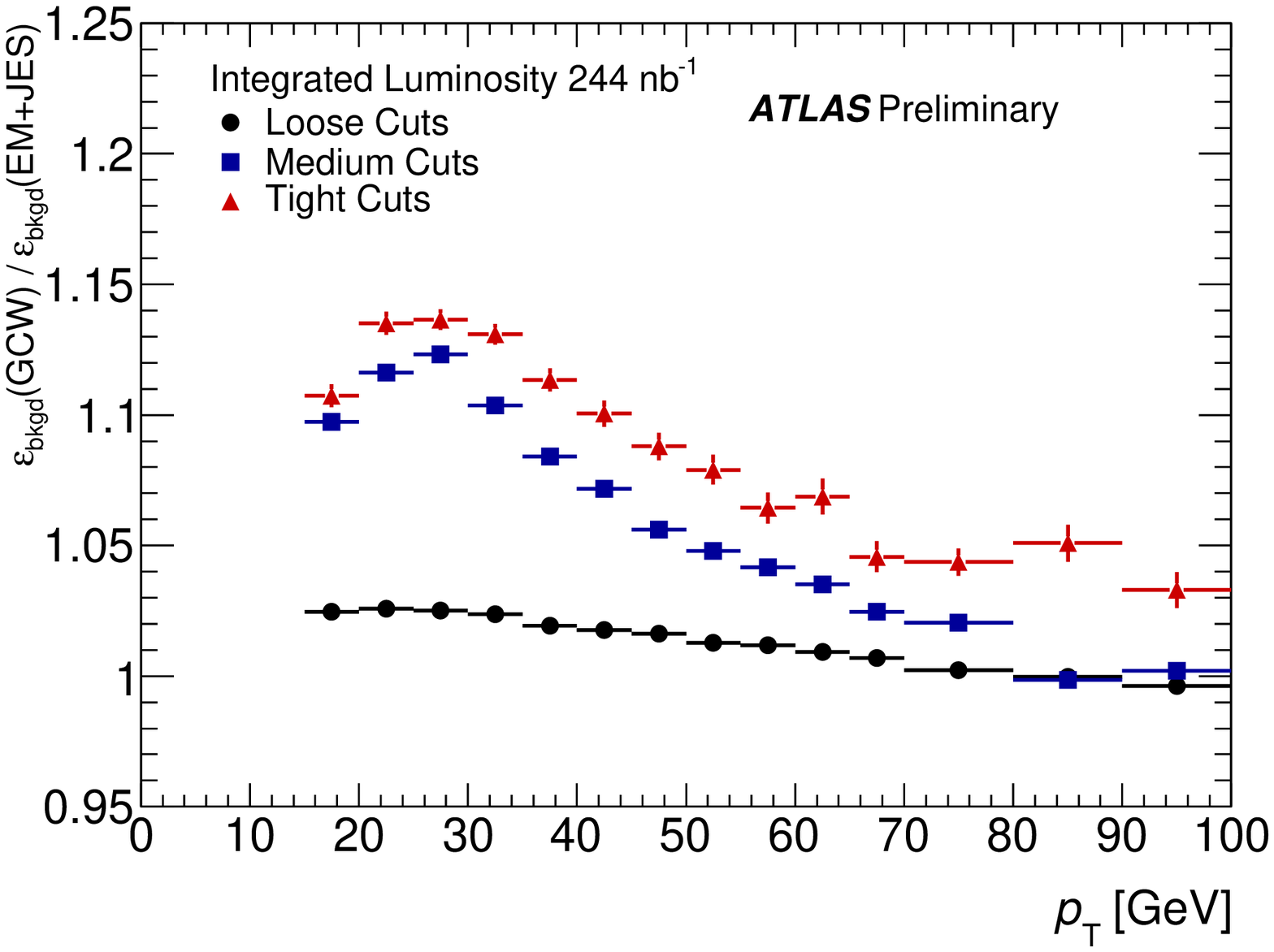}
  \label{subfig:calib}
}
\subfigure{
  \includegraphics[width=0.40\textwidth]{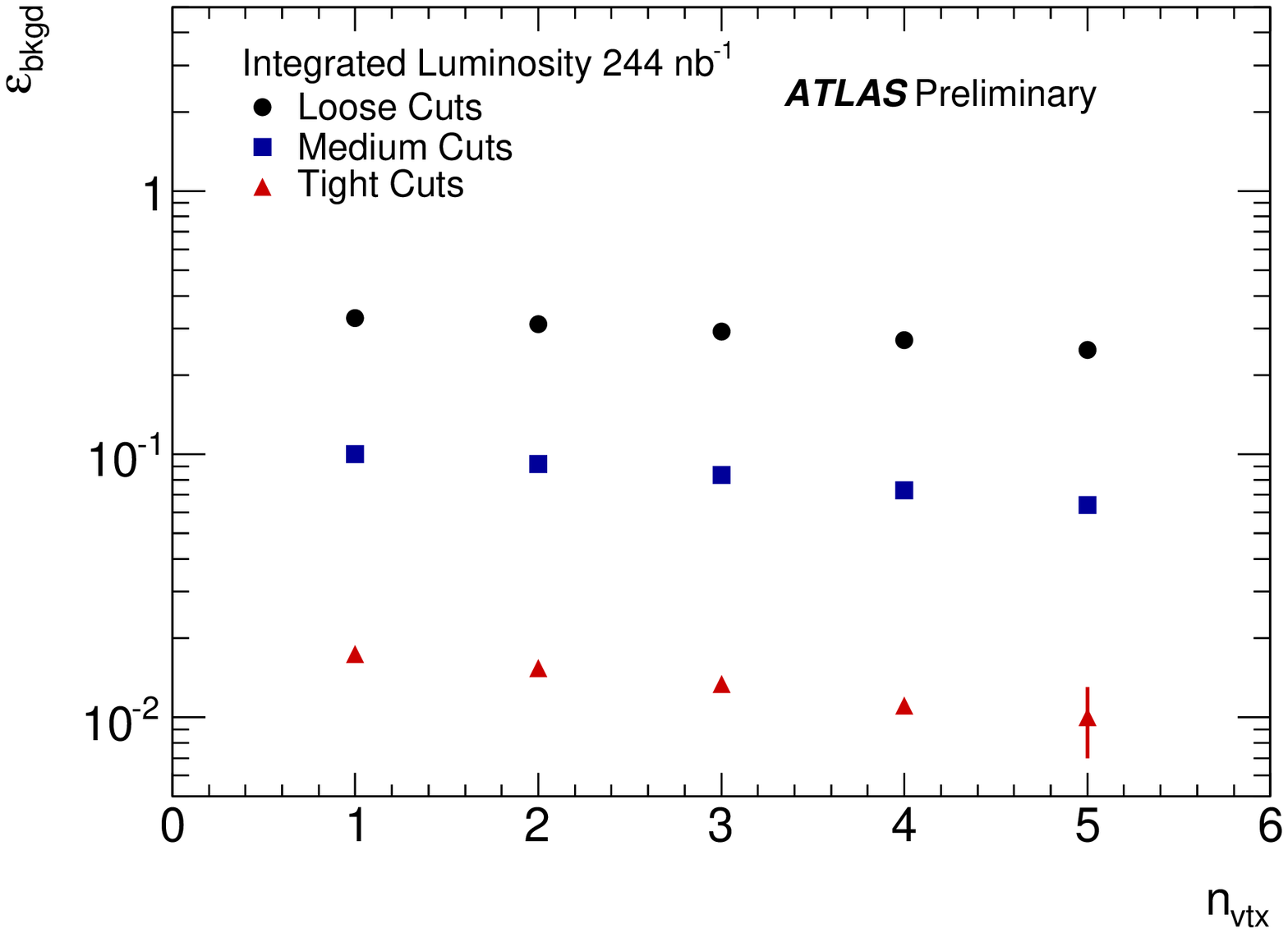}
  \label{subfig:eff}
}
\caption{Ratio of background efficiencies using EM+JES and GCW calibration as a function of $p_\mathrm{T}$ (left); background efficiencies as a function of $n_{vtx}$ (right) \protect\cite{conf086}.}
\label{fig:calib}
\end{figure}

Two effects contribute to systematic uncertainties:
\begin{itemize}
\item \textbf{The transverse momentum calibration:} Two calibration schemes have been compared, a global cell energy-density weighting (GCW) and a simple $p_\mathrm{T}$ and $\eta$ dependent calibration (EM+JES). The ratio of the background efficiency for both calibration schemes as function of $p_\mathrm{T}$ is shown in Fig.~\ref{fig:calib} (left).
\item \textbf{The pile-up effect:} During the data taking period, the beam intensity has increased significantly. The number of vertices $n_{vtx}$ is highly correlated with pile-up activity. The background efficiency as function of $n_{vtx}$ is shown in Fig.~\ref{fig:calib} (right).
\end{itemize}

\section{Conclusion}
All variables used in the $\tau$ ID algorithm are well described by MC predictions and show good separation power between $\tau$ leptons and fake $\tau$ candidates from QCD jets. Altogether the commissioning of the tauID was successful.

\begin{acknowledgments}
Work supported by the Helmholtz Young Investigators Group VH-NG-303.
\end{acknowledgments}

\end{document}